\pdfoutput=1

\documentclass[twocolumn, a4paper]{article}
\usepackage{graphicx}
\usepackage{amsmath, amsfonts, amssymb, amstext}
\usepackage{units}
\usepackage{textcomp}
\usepackage{gensymb}
\usepackage{wasysym}
\usepackage{cite}

\newcommand{\addPicture}[3] {
\begin{figure}[htb]
\begin{center}
\includegraphics[width=\columnwidth]{#1}
\caption{#2}
\label{#3}
\end{center}
\end{figure}
}

\title{
Silicon photomultiplier arrays -- a novel photon detector for a high resolution tracker produced at FBK-irst, Italy.
}

\author{
R.~Greim, H.~Gast, T.~Kirn, J.~Olzem, G.~Roper Yearwood, S.~Schael, N.~Zimmermann\\
\small{I. Physikalisches Institut B, RWTH Aachen University, 52074 Aachen, Germany} \smallskip \\
G.~Ambrosi, P.~Azzarello\\
\small{Dipartimento di Fisica, Universit\`a di Perugia, 06123  Perugia, Italy} \smallskip \\
R.~Battiston\\
\small{Dipartimento di Fisica, Universit\`a di Perugia, 06123  Perugia, Italy}\\
\small{Istituto Nazionale di Fisica Nucleare, Sezione di Perugia, 06123 Perugia, Italy} \smallskip \\
C.~Piemonte\\
\small{Fondazione Bruno Kessler - Istituto per la Ricerca Scientifica e tecnologica, 38050 Trento, Italy}
}

\date{}

\begin{document}

\maketitle

\begin{abstract}
A silicon photomultiplier (SiPM) array has been developed at FBK-irst \cite{piemonte} having 32 channels and a dimension of $\unit[8.0 \times 1.1]{mm^2}$. Each $\unit[250]{\micro m}$ wide channel is subdivided into $5 \times 22$ rectangularly arranged pixels. These sensors are developed to read out a modular high resolution scintillating fiber tracker. Key properties like breakdown voltage, gain and photon detection efficiency (PDE) are found to be homogeneous over all 32 channels of an SiPM array. This could make scintillating fiber trackers with SiPM array readout a promising alternative to available tracker technologies, if noise properties and the PDE are improved.
\end{abstract}

\section{Silicon Photomultipliers}

SiPMs are insensitive to magnetic fields, which make them capable of being used inside a charged-particle spectrometer. For that purpose, 32 prototypes of $\unit[8.0 \times 1.1]{mm^2}$ large SiPM arrays have been produced in 2007 at FBK-irst, Italy. Each array consists of 32 independent SiPMs, covering an area of $\unit[230 \times 1100]{\micro m^2}$ each with $5 \times 22$ pixels. The gap between the SiPM array channels is $\unit[20]{\micro m}$. The devices have a red/green sensitive n/p-doping scheme. A microscope picture of such a device can be seen in Fig.~\ref{fig:array}.

A silicon photomultiplier consists of an array of avalanche photodiodes (APD) pixels. Several APDs are connected in parallel together with a quenching resistor in series. The APDs are operated above their breakdown voltage. This operation mode leads to a high intrinsic gain of typically $G=10^5-10^6$ and to a dark count rate $f \approx \unit[100-5000]{\frac{kHz}{mm^2}}$ depending on the bias voltage and temperature.

A photon hitting the surface of an SiPM causes a pixel to break down with a probability $\epsilon_\mathrm{pde}$. The deposited charge is
$$Q = Ge = C \cdot \Delta U,$$
where $C$ is the capacitance of the pixel. The difference $\Delta U = U_\mathrm{bias}-U_\mathrm{0}$ between the bias and breakdown voltage is called overvoltage. During an avalanche process photons are produced, that can cause adjacent pixels to break down with a probability $p_\mathrm{xtalk}$ \cite{otte}. This effect leads to a nonlinear excess of the signal.

Increasing $\Delta U$ has the positive effect of a higher $G$ and $\epsilon_\mathrm{pde}$, but unfortunately also increases the dark count rate and $p_\mathrm{xtalk}$. These values have to be balanced to determine the operation voltage.

\section{Tracker Module Design}

A modular high resolution scintillating fiber tracker with SiPM array readout has been developed for the PEBS \cite{doetinchem} experiment. The module comprises two stacks of round scintillating fibers mounted to a support structure. Each stack consists of $5 \times 128$ fibers with a diameter of $\unit[250]{\micro m}$ in the tightest arrangement. On both sides of the module, a hybrid with four SiPM arrays is fastened to the support structure. One end of the fibers is covered by a mirror for a better light yield. A schematic drawing of a tracker module can be seen in fig.~\ref{fig:trackerModule}.

In a testbeam in 2008 the mean number of detected photons was measured to be on average 5 per minimum ionizing particle for the combination of Kuraray SCSF-81M fibers and FBK-irst 2007 SiPM arrays \cite{roper}.

\section{I-V-Characteristic Curves}

To determine the operation voltage of the SiPM array, the flowing current is measured as a function of the applied reverse bias voltage. The properties of an SiPM strongly depend on the applied overvoltage, thus on the breakdown voltage $U_0$. Variations in $U_0$ lead to variations in key properties like $p_\mathrm{xtalk}$, $G$ and $\epsilon_\mathrm{pde}$.

In Fig.~\ref{fig:iv} the I-V-characteristic curves of all 32 channels are plotted as a function of the bias voltage for one SiPM array. A function
$$I\left(U\right) = \left\{ \begin{array}{ll} I_0 & U < U_0 \\ I_0 + L \left(U-U_0\right) & \text{otherwise} \end{array} \right.$$
is fitted to each curve to determine $U_0$. The average of $U_0$ is $\unit[30.1]{V}$ with a variation of $\unit[0.4]{\%}$. This shows the high homogeneity of the device. 

\section{Gain}
A homogeneous gain over the 32 channels of an SiPM array simplifies the detector operation. SiPM spectra have been taken with a pulsed LED (Fig.~\ref{fig:gainCompare}). For this measurement we used an VA\_32/75 based readout chip hybrid connected to the DAQ system of the AMS02 silicon tracker. The charge deposit is plotted versus each SiPM array channel.

The pedestal and the photoelectron peaks corresponding to the number of fired pixels can clearly be seen. Additionally, the center of gravity of the first peaks are shown (dots). The distance between the photoelectron peaks corresponds the gain $G$. The average gain is 119.2 ADC counts with a variation of $\unit[2]{\%}$, again showing the homogeneous response.

\section{Fill Factor}
The fill factor or geometric efficiency is defined as the ratio of sensitive to total area of a pixel. A test setup has been built to determine the fill factor.

In the setup an SiPM is located under the lens of a microscope and can be moved using two crossed motorized linear stages. The microscope is used to focus an LED spot to a diameter of less than $\unit[5]{\micro m}$. The surface of an SiPM is scanned two-dimensionally with a step width of $\unit[1]{\micro m}$ shooting 50 short LED pulses at each position. If the SiPM breaks down, a point of the SiPM surface belongs to the sensitive area. The ratio of sensitive to the total number of points is then the fill factor of a pixel.

In Fig.~\ref{fig:fillFactor}, the results for nine pixels of an FBK-irst 2007 SiPM array are shown. On average, a fill factor $\epsilon_\mathrm{ff} = \unit[\left(43.9 \pm 0.6_\mathrm{rms}\right)]{\%}$ has been found. There is a rather large gap between the sensitive area of the pixels, which has the advantage of a crosstalk probability of only $\sim \unit[10]{\%}$, but also causes a small $\epsilon_\mathrm{ff}$.

\section{Photon Detection Efficiency}
The photon detection efficiency $\epsilon_\mathrm{pde}$ is defined as the product
$$\epsilon_\mathrm{pde} = \epsilon_\mathrm{qe} \cdot \epsilon_\mathrm{ff} \cdot \epsilon_\mathrm{abe},$$
where $\epsilon_\mathrm{qe}$ is the quantum efficiency of the APD pixels, $\epsilon_\mathrm{ff}$ the fill factor and $\epsilon_\mathrm{abe}$ the avalanche breakdown efficiency. With an increasing overvoltage $\epsilon_\mathrm{abe}$ increases, thus leading to a higher $\epsilon_\mathrm{pde}$. The dependence is shown in Fig.~\ref{fig:pdeBias}. At $U_\mathrm{bias} = \unit[32.5]{V}$, $\epsilon_\mathrm{pde} = \unit[\left(22.3 \pm 0.6_\mathrm{rms}\right)]{\%} $ is reached.

An $\epsilon_\mathrm{pde} \approx \unit[50]{\%}$ for SiPMs with comparable pixel size and $\epsilon_\mathrm{ff} = \unit[61.5]{\%}$ has already been achieved at a wavelength of \unit[400]{nm} \cite{hamamatsu}. The lower $\epsilon_\mathrm{pde}$ is mainly due to the smaller geometric fill factor of FBK-irst 2007 SiPM arrays. A new array has been produced at FBK-irst in 2008 with a two times higher fill factor, optimized for fiber trackers. According to calculations, $\epsilon_\mathrm{pde}$ of the new FBK-irst array is expected to match the results achieved in \cite{hamamatsu}.

\section{Noise}
In a tracker a particle passage is identified by a charge deposit in an SiPM array channel exceeding a certain threshold. Due to the operation above the breakdown voltage, SiPM pixels can fire thermically, which causes fake hits inside the tracker. To determine the fake hit probability, dark spectra of the SiPM array channels have been recorded at a bias voltage of $\unit[32.5]{V}$ and a temperature $T = \unit[23]{\celsius}$. In Fig.~\ref{fig:noise}, a mean spectrum over the 32 SiPM array channels can be seen together with the fake hit probabilities at the according cut.

At an 0.5 photoelectron cut and a sampling window of $\unit[75]{ns}$, $\unit[14.5]{\%}$ of the detected hits in an SiPM array channels are fake hits. This probability decreases exponentially with the photoelectron cut. However, in order to achieve an acceptable tracking efficiency, the cut has to be set to $\unit[0.5]{photoelectrons}$.

\section{Conclusion}
The production of the first 32 large dimension SiPM arrays at FBK-irst in 2007 has been a success concerning the homogeneity of the device. To meet the requirements for the use in a high resolution scintillating fiber tracker the photon detection efficiency has to be improved. This can be done by increasing the fill factor. New arrays with more than two times higher fill factor have been produced at FBK-irst in 2008 and are currently under test. They
are expected to deliver at least a factor two higher photon detection efficiency. If in addition the issue of noise is reduced, e.\,g. by cooling the devices, a high resolution fiber tracker based on this technology is feasible.

\addPicture{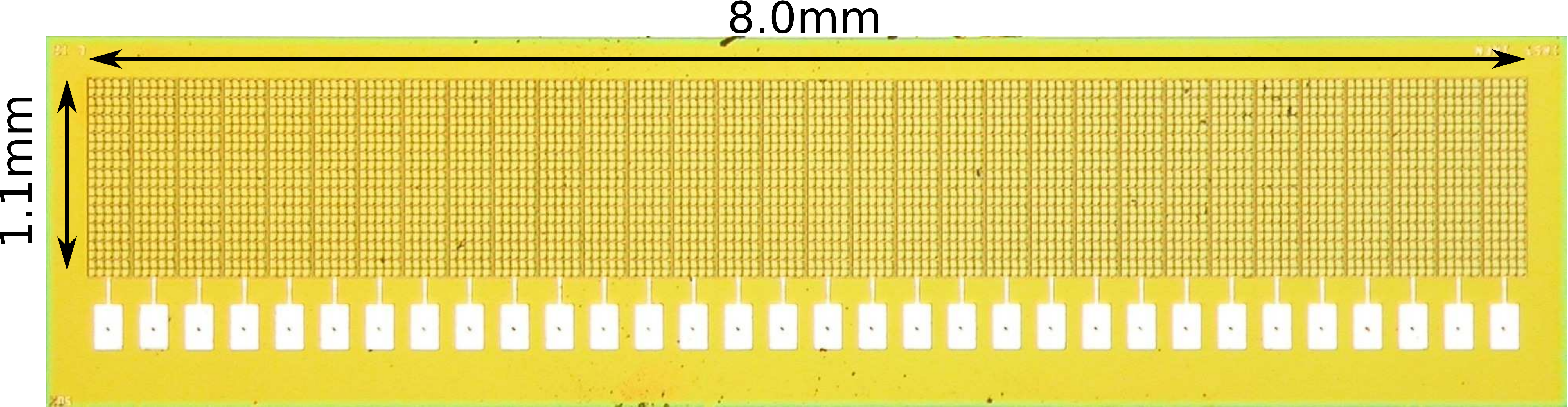}{FBK-irst 2007 SiPM array.}{fig:array}
\addPicture{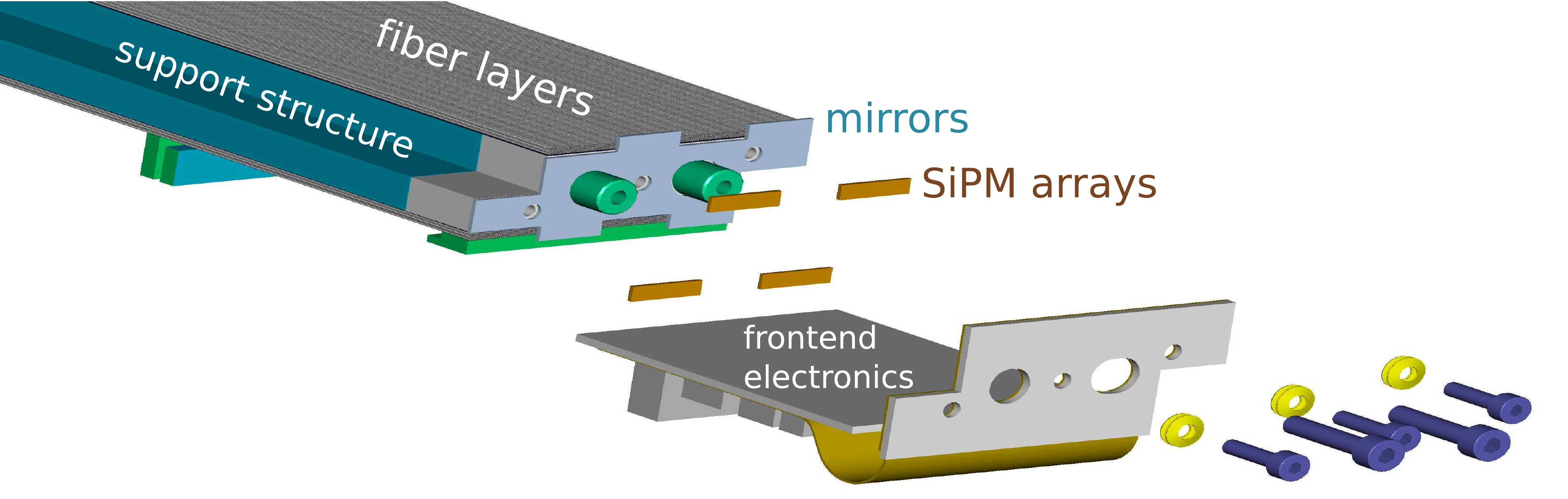}{Exploded view of a tracker module.}{fig:trackerModule}
\addPicture{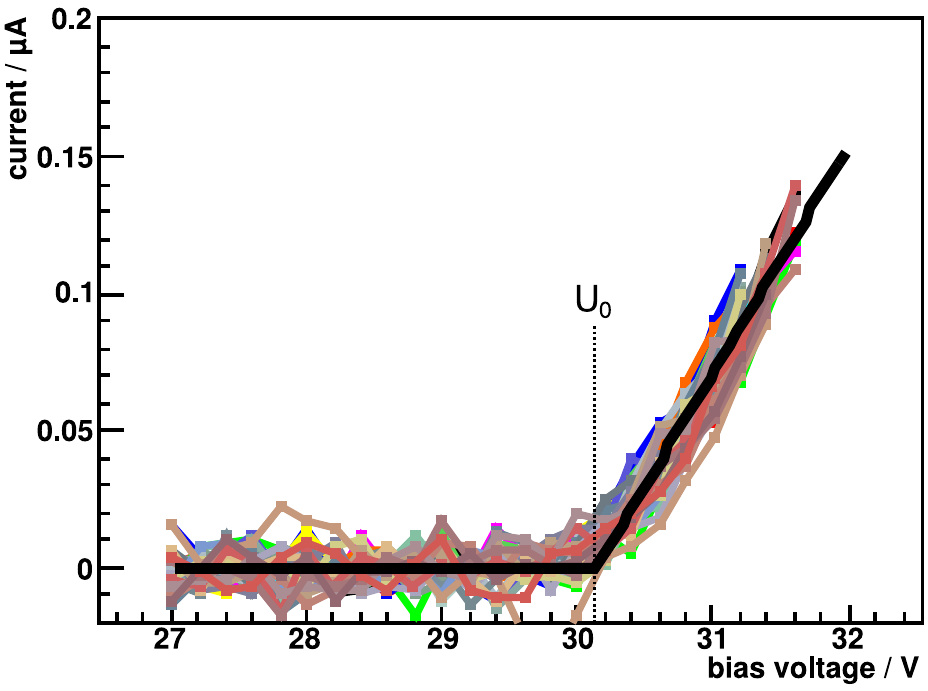}{I-V-characteristic curves of the 32 channels of an FBK-irst 2007 SiPM array. Additionally, the mean $I\left(U\right)$ is plotted in black.}{fig:iv}
\addPicture{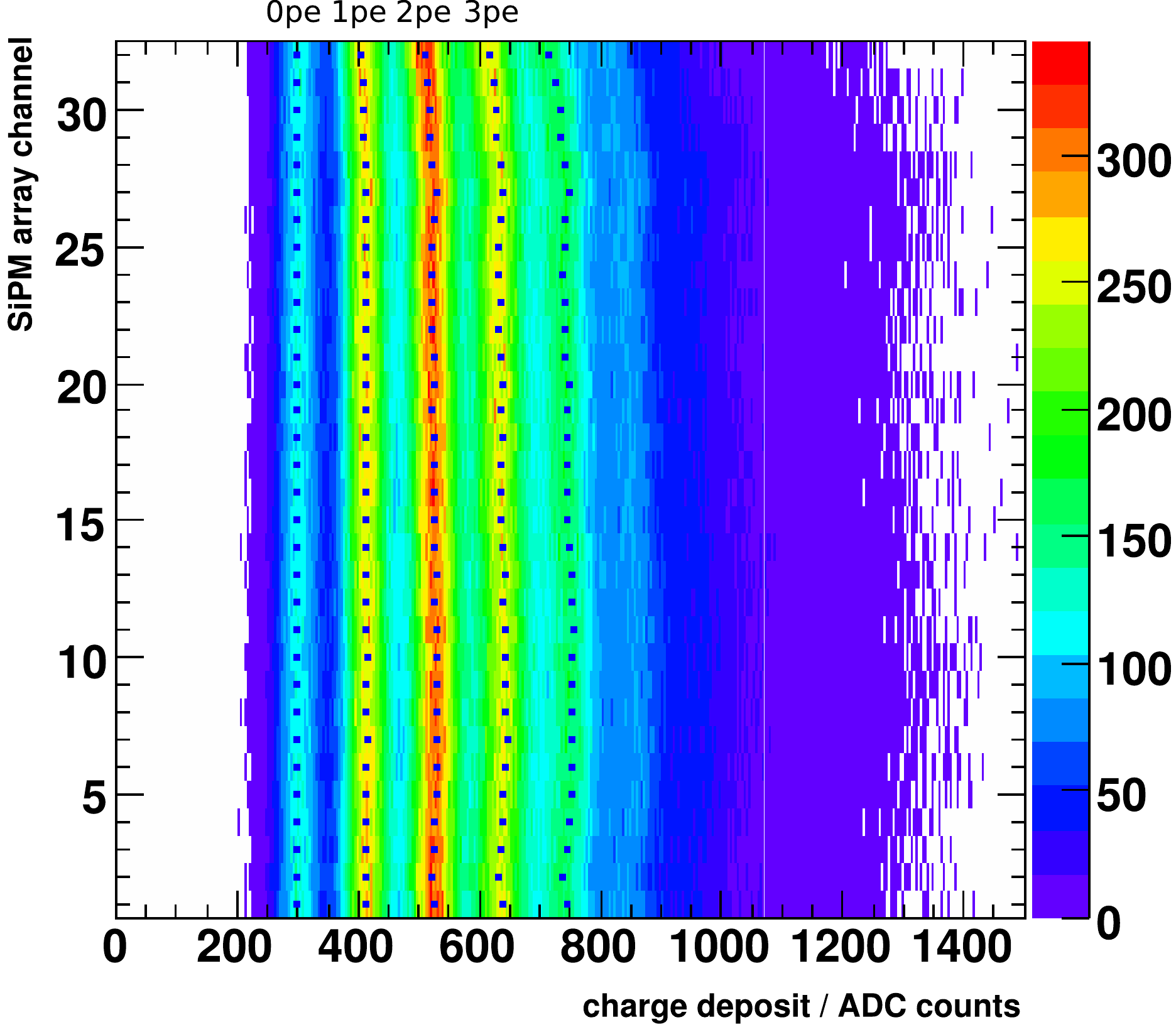}{Channelwise comparison of the 32 SiPM spectra of an FBK-irst 2007 SiPM array taken with an LED at $U_\mathrm{bias} = \unit[32.5]{V}$.}{fig:gainCompare}
\addPicture{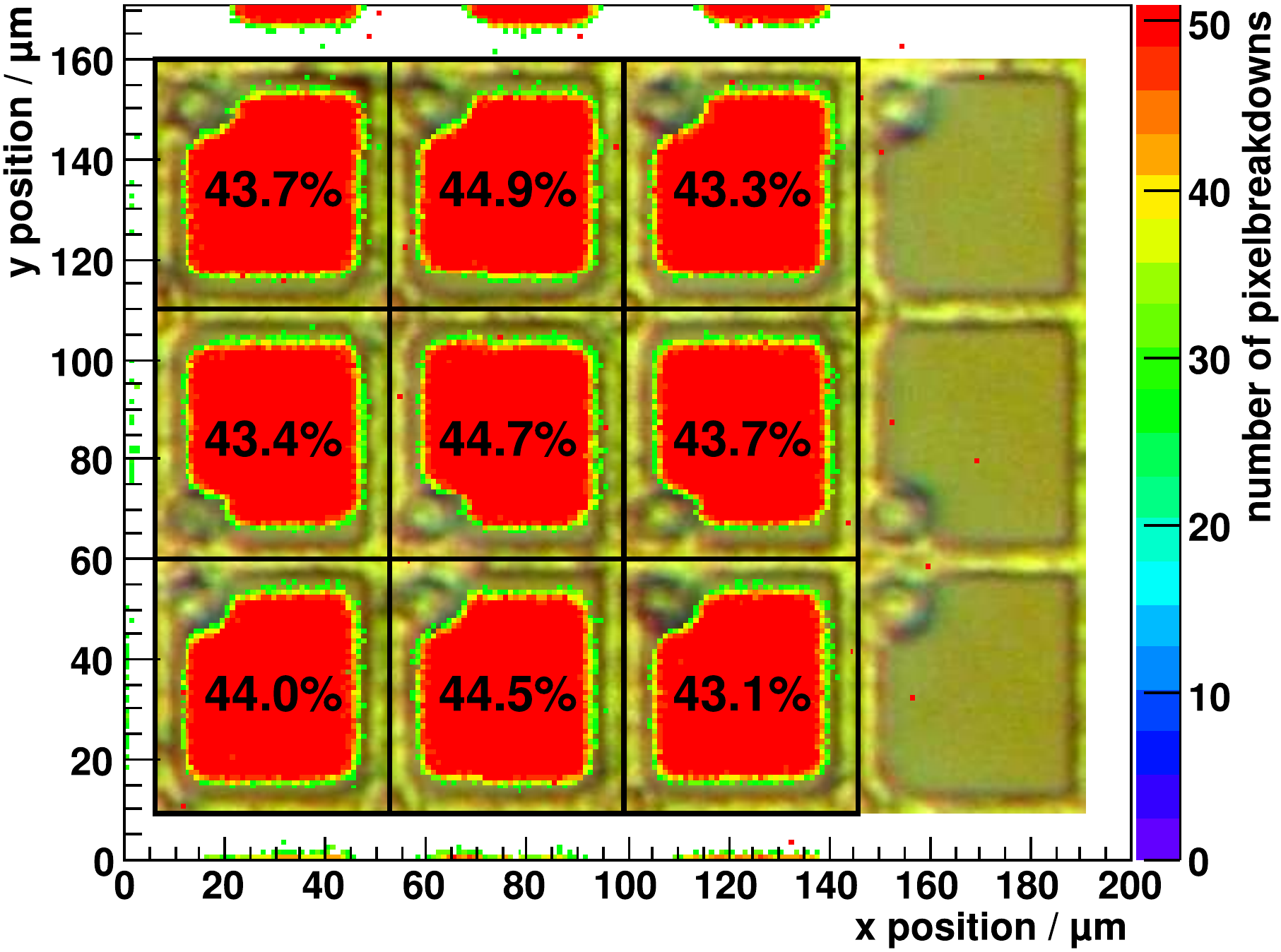}{Fill factor of nine pixels of an FBK-irst 2007 SiPM array. Additionally, a microscope picture of the surface is shown. Entries with a low frequency have been removed.}{fig:fillFactor}
\addPicture{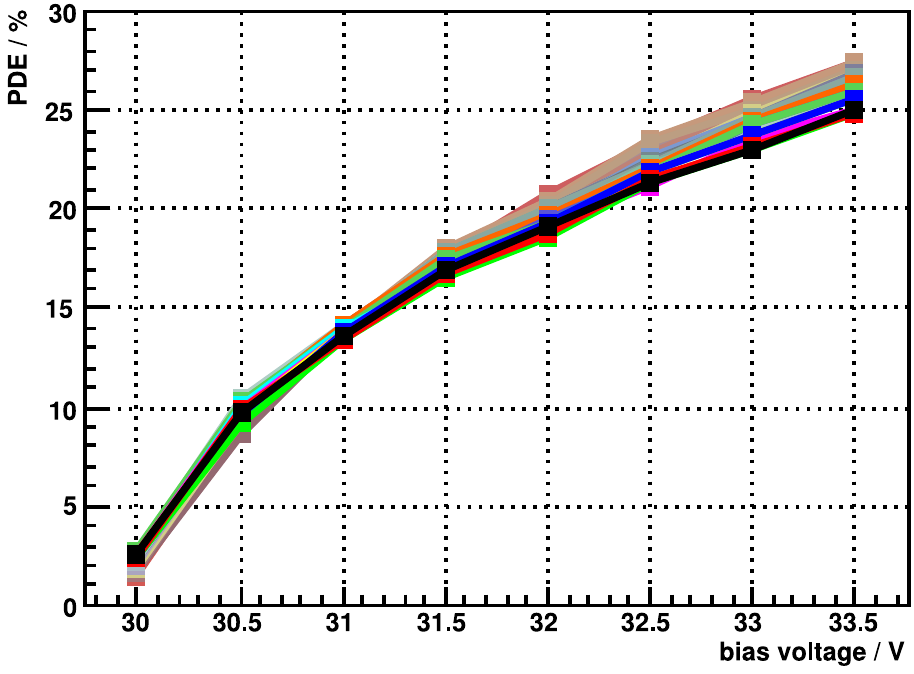}{Dependence of the photon detection efficiency on the bias voltage at a wavelength of \unit[495]{nm}.}{fig:pdeBias}
\addPicture{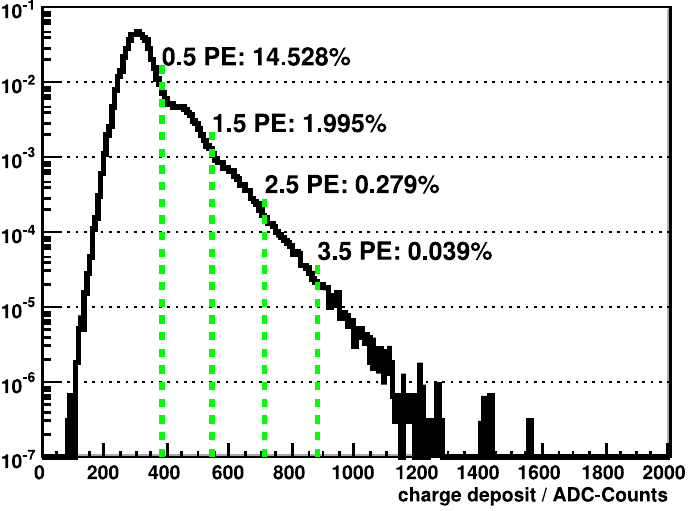}{Fake hit probability depending on the applied photoelectron cut at $U_\mathrm{bias} = \unit[32.5]{V}$ and $T = \unit[23]{\celsius}$.}{fig:noise}

\end{document}